# High Sensitivity and Specificity Biomechanical Imaging by Stimulated Brillouin Scattering Microscopy

Itay Remer[1,2], Netta Shemsesh[3], Anat Ben-Zvi[3] & Alberto Bilenca[1,4*]

[1]Biomedical Engineering Department, Ben-Gurion University of the Negev, 1 Ben Gurion Blvd, Be'er-Sheva 84105, Israel
[2]Agilent Research Laboratories, 94 Shlomo Shmeltzer Road, Petach Tikva 4970602, Israel
[3]Department of Life Sciences, Ben-Gurion University of the Negev, 1 Ben Gurion Blvd, Be'er Sheva 84105, Israel
[4]Ilse Katz Institute for Nanoscale Science and Technology, Ben-Gurion University of the Negev,
1 Ben Gurion Blvd, Be'er-Sheva 84105, Israel

e-mail: *bilenca@bgu.ac.il

## Abstract

Noncontact label-free biomechanical imaging is a crucial tool for unraveling the mechanical properties of biological systems, which play critical roles in the fields of engineering, physics, biology and medicine; yet, it represents a significant challenge in microscopy. Spontaneous Brillouin microscopy meets this challenge, but often requires long acquisition times or lacks high specificity for detecting biomechanical constituents with highly overlapping Brillouin bands. We developed stimulated Brillouin scattering (SBS) microscopy that provides intrinsic noncontact biomechanical contrast and generates mechanical cross-sectional images inside large specimens, with high mechanical specificity and pixel dwell times that are >10-fold improved over those of spontaneous Brillouin microscopy. We used SBS microscopy in different biological applications, including the quantification of the high-frequency complex longitudinal modulus of the pharyngeal region of live wild-type *Caenorhabditis elegans* nematodes, imaging of the variations in the high-frequency viscoelastic response to osmotic stress in the head of living worms, and in vivo mechanical contrast mesoscopy of developing nematodes.

## Main Text

Label-free biomechanical imaging has long used a variety of techniques, including atomic-force microscopy, multiphoton microscopy, and optical coherence elastography[1–6], that obtain mechanical images with high spatial resolution, but require contact or external mechanical stimulation of the sample. Spontaneous Brillouin microscopy[7–20] circumvents these requirements by measuring the so-called Brillouin shifts $\Omega_B$ and linewidths $\Gamma_B$, which are the frequency shifts and linewidths of light backscattered inelastically from gigahertz-frequency longitudinal acoustic phonons characteristic to the different viscoelastic constituents of the material. However, spontaneous Brillouin microscopy often demands long acquisition times due to the low efficiency of spontaneous Brillouin scattering in biological matter, or suffers from limited mechanical specificity because of the relatively low spectrometer resolution of spontaneous Brillouin microscopes, making it difficult to specifically detect biomechanical constituents with highly overlapping Brillouin bands.

Here we introduce stimulated Brillouin scattering (SBS) as a process to significantly enhance the acquisition rate and mechanical specificity of Brillouin microscopy. SBS is a photon-phonon scattering process analogous to stimulated Raman scattering but which involves lower-frequency acoustic phonons. SBS was first observed in 1964[21] and has widely been used in spectroscopic research[22–29]. While in spontaneous Brillouin scattering a single laser beam at a frequency $\omega_1$ illuminates the sample to produce the entire Brillouin spectrum around the Stokes and anti-Stokes frequencies $\omega_1 \pm \Omega_B$, in SBS counter narrowband laser pump and probe beams at $\omega_1$ and $\omega_2$ overlap inside the sample to efficiently interact with a particular longitudinal acoustic phonon of vibrational frequency $\Omega_B$ (Fig. 1a). When the frequency of the probe beam at $\omega_2$ is scanned around the Stokes frequency $\omega_1 - \Omega_B$, its intensity $I_2^S$ experiences a stimulated Brillouin gain (SBG) via wave resonance, whereas $I_1$ the intensity of the pump beam at $\omega_1$ shows a stimulated Brillouin loss (SBL), as depicted in Fig. 1b. The opposite occurs when $\omega_2$ is scanned around the anti-Stokes frequency $\omega_1 + \Omega_B$. No gain or loss arise at $\omega_2$ otherwise. Consequently, unlike spontaneous Brillouin scattering, SBG or SBL enables Brillouin spectrum measurements that are free of elastic background interference and that do not trade spectrometer resolution for acquisition time, or vice versa[25–28].

The SBG or SBL spectrum is given by $G(\Omega) = \pm \eta \times g(\Omega) \times \ell \times I_1$, where $\eta$ is the crossing efficiency of the pump and probe beams in the sample, $\pm g(\Omega)$ is the SBS gain or loss factor well represented by a lorentzian line shape, $\ell$ is the interaction length of the two counter laser beams in the sampled volume, and $I_1$ is the intensity of the pump beam at $\omega_1$[23]. As for the spontaneous Brillouin scattering spectrum[7, 10], the close relationship between the SBG or SBL spectrum and the complex longitudinal modulus of the probed volume allows to locally extract the high-frequency viscoelastic response of the medium. This relationship is described by $M^* = \rho \times (\lambda_1/2n)^2 \times \Omega_B^2 \times (1 + i\Gamma_B/\Omega_B)$, where $M^*$ is the complex longitudinal modulus, $\lambda_1$ is the wavelength of the beam at $\omega_1$, and $n$ and $\rho$ are the refractive index and the mass density of the medium, respectively[7, 30]. Similar to other nonlinear optical techniques[31], SBS inherently provides optical sectioning in three dimensions owing to the nonlinearity of SBS in the total irradiance intensity.

SBS has recently been demonstrated for Brillouin microscopy of liquids using tunable single-frequency lasers in a frequency-scanning SBS scheme[24] or picosecond lasers in an impulsive SBS arrangement[29]. Although these systems provide high spectrometer resolution, their low sensitivity hinders biomechanical imaging.

To enable biomechanical SBS imaging, we constructed a SBS microscope (Fig. 1c; see Methods Section 1 and Supplementary Fig. 1) based on our previous frequency-scanning SBS spectrometer design[26–28], but with an improved spatial resolution of ~0.8 × 0.8 × 16 μm³ and an enhanced shot-noise sensitivity that yields a spectral acquisition time as low as 2 ms in water with Brillouin shift and linewidth measurement precision of 11.5 MHz and 35 MHz, respectively (see Methods Section 2 and Supplementary Fig. 2). This acquisition time represents a ≥50-fold improvement over the typical spectral measurement times of hundreds of milliseconds used in Brillouin measurements of water[11-13,16–19]. The obtainable Brillouin shift and linewidth measurement precision allow to determine the real and imaginary parts of the gigahertz-frequency longitudinal modulus of water to a fractional precision of 0.0046 and 0.11, respectively (see Methods Section 2). The light sources were 780-nm tunable distributed-feedback semiconductor lasers that delivered a total excitation power of ~265 mW at the sample. The linewidth of the lasers was 1-4 MHz, enabling highly specific measurements of a SBG or SBL spectrum with megahertz spectrometer resolution, which is at least 12-fold higher than the >100-MHz spectrometer resolution of Brillouin biomicroscopes[9–20].

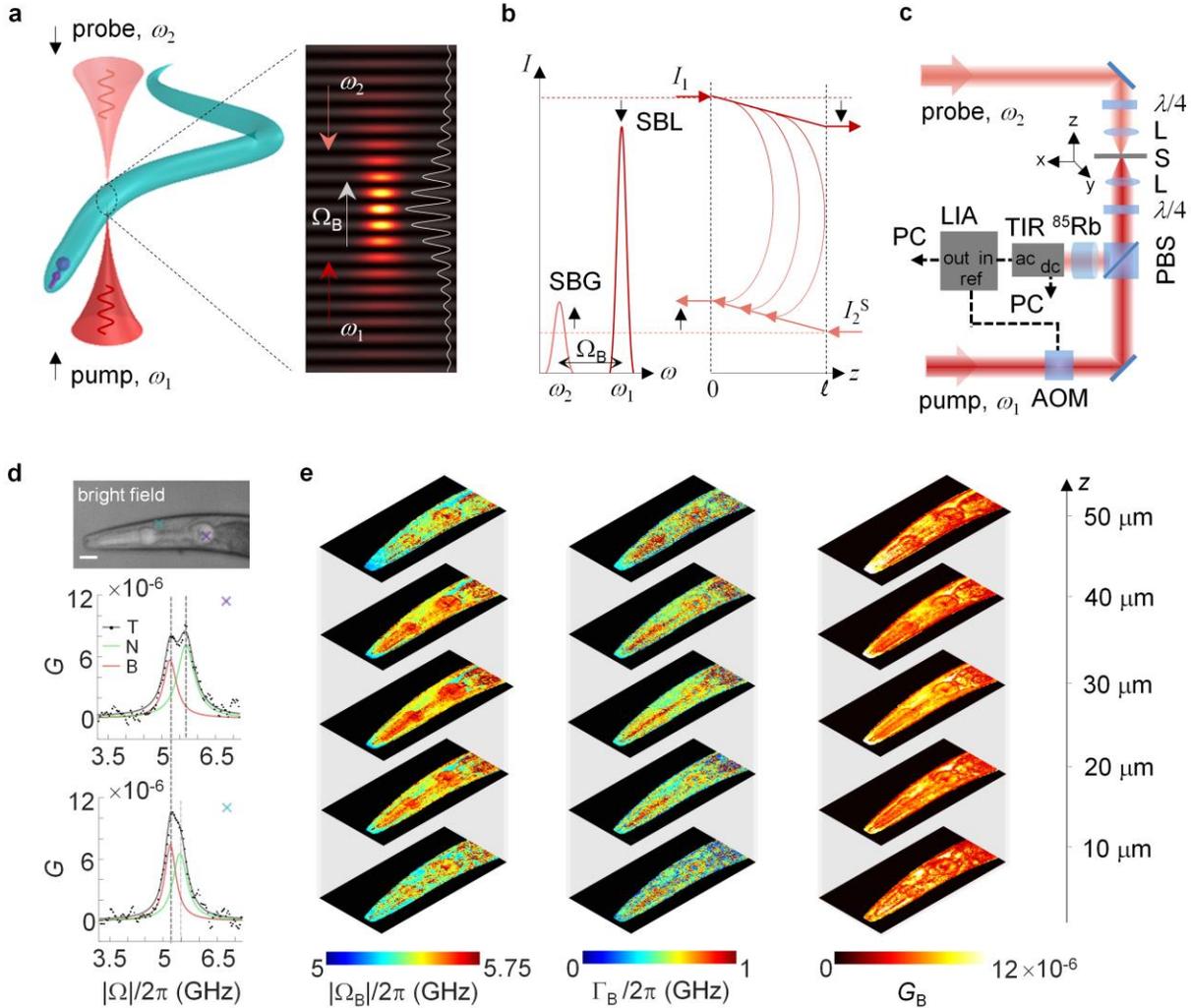

**Fig. 1.** Principle and method of SBS microscopy. (**a**) Acousto-optic interaction between focused, counter-propagating pump ($\omega_1$) and probe ($\omega_2 < \omega_1$) light with longitudinal acoustic phonon ($\Omega_B$). The interference of the pump and probe fields (stripe pattern with intensity profile in white) drives the acoustic wave (gray wavefronts). (**b**) Intensity transfer between the pump ($I_1$) and the probe ($I_2^S$) light by virtue of SBS along an interaction length $\ell$, with an intensity increase in the probe beam (SBG) and an intensity decrease in the pump beam (SBL). Left and right illustrations are in the frequency and spatial domains, respectively. (**c**) Schematic of the SBS microscope. The pump beam is modulated by an acousto-optic modulator (AOM) and is left circularly polarized by a quarter wave plate ($\lambda/4$). The probe beam is right circularly polarized by a second quarter wave plate ($\lambda/4$). The pump and the probe beams are focused to the same point in the sample (S) by identical lenses (L). The frequency of the probe beam is scanned around the Stokes frequency $\omega_1 - \Omega_B$ to produce the backscattering SBG spectrum of the focus point on the probe intensity. The probe beam is directed to an atomic notch filter (AF) by a polarizing beam splitter (PBS). The filter passes the probe light to a custom transimpedance receiver (TIR) and suppresses the unwanted pump back-reflections into the receiver. The ac signal at the receiver is input to a lock-in amplifier (LIA) to measure the SBG spectrum, whereas the dc signal is processed by a personal computer (PC) to measure the total attenuation across the sample. By raster-scanning the sample through the mutual focus of the pump and probe beams, three-dimensional images are obtained. (**d**) SBG

spectrum from volumes inside the pharynx (upper spectrum; purple cross in the coregistered brightfield image, scale bar = 20 μm) and the surrounding region (bottom spectrum; cyan cross in the coregistered brightfield image) of a live wild-type *C. elegans* nematode. The spectra measured (T, dots) in 20 ms show a multipeak feature that was fitted well by a double lorentzian line shape (T, black line), with contributions from a particular mechanical constituent of the nematode (N, green line) and the buffer component (B, red line). (**e**) Three-dimensional sections of the Brillouin shift $\Omega_B$, Brillouin linewidth $\Gamma_B$, and peak SBG $G_B$, through the head of a live *C. elegans* nematode. A pixel dwell time of 20 ms was used to acquire a SBG spectrum over the range of 3-7 GHz in each pixel. Images contain 100 × 200 pixels, resulting in a total image recording time of 400 s.

We show in Fig. 1d the SBG spectrum from volumes inside the pharynx and the surrounding region of a live wild-type *C. elegans* nematode (see Methods Sections 1 and 3, and Supplementary Fig. 3), which is an important model organism for biomedical research[32]. No substantial photodamage was observed (see Methods Section 4 and Supplementary Fig. 4). A multipeak feature is apparent in these two spectra, which can be well reproduced by a double lorentzian line shape. Whereas the green lorentzian represents the SBG spectral contribution of a particular mechanical constituent of the nematode, and the red lorentzian corresponds to the SBG spectral contribution of the buffer component, which always exists in the sampled volume. Thus, SBS enables highly specific identification of the Brillouin shift $\Omega_B$, linewidth $\Gamma_B$, and peak gain $G_B$ that completely parametrize the SBG spectrum of a specific biomechanical component of the specimen. By raster scanning the sample through the focus of the microscope, cross-sectional images of $\Omega_B$, $\Gamma_B$, and $G_B$ of the nematode were obtained across the entire 50 μm thickness of the worm's head, as shown in Fig. 1e.

As the first application of SBS microscopy, we imaged and quantified the gigahertz-frequency complex longitudinal modulus in live wild-type *C. elegans* at the second larval stage (L2) (Fig. 2). Figure 2a displays the Brillouin shift $\Omega_B$ image and the coregistered brightfield image of a live *C. elegans* larva along with images of the Brillouin linewidth $\Gamma_B$ (Fig. 2b) and peak gain $G_B$ (Fig. 2c) across the entire middle plane of the worm at a depth of 10 μm. The pharynx, analogous to the vertebrate esophagus, is an important organ of the nematode that pumps and grinds food into the worm's intestine[33]. The pharynx is seen to exhibit high values of Brillouin shift and linewidth but low values of peak gain compared to the surrounding region. From the SBG spectral parameters ($\Omega_B$, $\Gamma_B$, $G_B$) measured in the worm, and using the relation between mass density and SBG[23] and the reported refractive index distribution of the worm[34], the spatial distributions of the gigahertz-frequency complex longitudinal moduli in the nematode were evaluated, as shown in Fig. 2d,e (see Methods Sections 5 and 6, and Supplementary Fig. 5). These distributions indicate that at high frequencies the larva pharynx is stiffer—with higher Brillouin shifts (Fig. 2a)—and more viscous—with larger Brillouin linewidths (Fig. 2b)—than its surroundings, attributable to the muscular nature of the pharynx[33]. Such direct correlations between the Brillouin shift and stiffness and the Brillouin linewidth and viscosity were consistently observed in other highly hydrated biological systems[10, 17]. We further confirmed these findings by statistically comparing the Brillouin shift and linewidth as well as the complex longitudinal modulus of the pharynx (purple in Fig. 2a) and the surrounding region (cyan in Fig. 2a) averaged across the corresponding volumes in several worms (Fig 2f-i). Therefore, we can use SBS microscopy to map the gigahertz-frequency complex longitudinal modulus of live organisms, opening possibilities to study biomechanics at high frequencies.

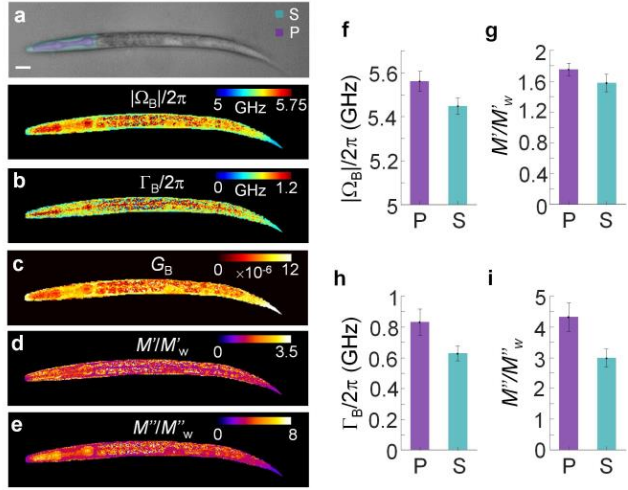

**Fig. 2.** Characterization of the high-frequency complex longitudinal modulus in live wild-type *C. elegans* L2 larvae with SBS microscopy. (**a**) Brillouin shift $\Omega_B$ image (and the coregistered brightfield image) of a live L2 larva at a depth of 10 μm inside the worm. In the brightfield image, the pharynx (P) and the surrounding region (S) are filled in purple and cyan, respectively. The pixel dwell time and the frequency acquisition range were as in Fig. 1e. Images contain 100 × 400 pixels, resulting in a total image recording time of 800 s. Scale bar, 20 μm. (**b**) Brillouin linewidth $\Gamma_B$ image of the larva. (**c**) Peak SBG $G_B$ image of the larva. (**d**) Gigahertz-frequency longitudinal storage modulus $M'$ of the larva (relative to that of double-distilled water $M'_W = 2.056$ GPa). Mean refractive index values from the literature were used (34) for the pharynx ($n = 1.38$) and the surrounding region ($n = 1.36$). For the other regions of the worm, a mean refractive index value of 1.37 was used[34]. (**e**) Gigahertz-frequency longitudinal loss modulus $M''$ of the larva (relative to that of double-distilled water $M''_W = 0.1249$ GPa). (**f**) Average Brillouin shift of the pharynx (P, purple bar) and the surroundings (S, cyan bar) of live L2 larvae ($N = 10$, $p$-value $< 0.05$ as computed by paired student's $t$-test, 10-μm depth). Error bars represent standard deviation from the mean. (**g**) Relative average gigahertz-frequency longitudinal storage modulus $M'$ of the larva pharynx and the surrounding region (same statistics as in **f**). (**h**) Average Brillouin linewidth of the larva pharynx and surroundings (same statistics as in **f**). (**i**) Relative average gigahertz-frequency longitudinal loss modulus $M''$ of the larva pharynx and surroundings (same statistics as in **f**). These results indicate that at gigahertz frequencies the larva pharynx is stiffer and more viscous than its surroundings.

We also present the use of SBS microscopy to visualize variations in the high-frequency viscoelastic response to hyperosmotic stress within the pharyngeal region of live wild-type *C. elegans* young adults. Common instruments for biomechanics, such as cantilevers and micropipettes, are limited to measurements at the surface of the worm and cannot image biomechanical properties in vivo with depth sectioning[2, 35–38]. Figure 3A-C shows the brightfield images and the Brillouin shift $\Omega_B$ and linewidth $\Gamma_B$ sections of the head region of three individual *C. elegans* young adults from a depth of 25 μm, under isotonic (Fig. 3a) and hypertonic (Fig. 3b,c) salt conditions. At 125 mM of salt (250 Osm), the images show a notable increase in the Brillouin shift and linewidth throughout the worm's head compared to a control worm under isotonic conditions. At 250 mM of salt (500 Osm), the Brillouin shift and linewidth further increased. These results suggest that hyperosmotic shock increases the high-frequency stiffness and viscosity of the worm's pharyngeal region; consistent with previous findings that exposure of *C. elegans* to a high-salt environment induces an increase in the low-frequency stiffness of the animal's body-wall[37]. Similar results were reported in biological cells using spontaneous Brillouin microscopy and atomic-force microscopy[10]. At both isotonic and hypertonic salt concentrations, the Brillouin shift and linewidth of the worm's pharynx are found to remain higher than those of the surrounding area, indicating that the pharynx is stiffer and more viscous than its surroundings regardless of the osmotic conditions. All these findings were further verified statistically in several nematodes by means of the average Brillouin shift and linewidth of the pharynx and the surrounding region (Fig. 3d,e). These results show that SBS microscopy offers a new approach for investigating biomechanics in vivo.

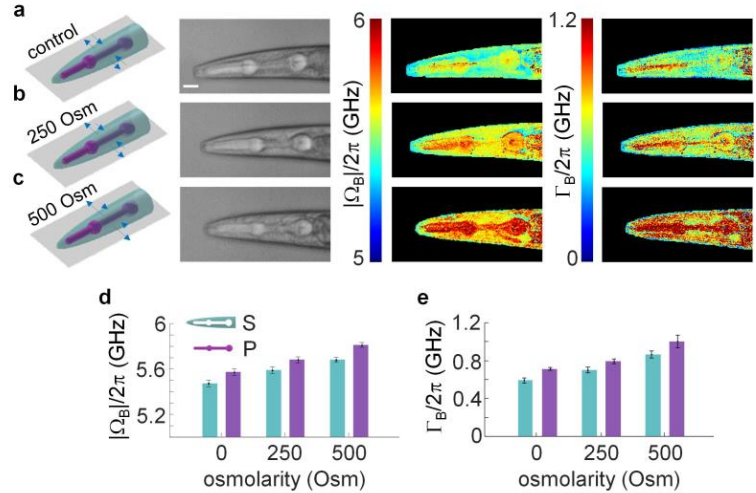

**Fig. 3.** SBS imaging of the variations in the high-frequency viscoelastic response to hyperosmotic stress within the pharyngeal region of live wild-type *C. elegans* young adults. (**a-c**) Brillouin shift $\Omega_B$ and Brillouin linewidth $\Gamma_B$ images (and the coregistered brightfield image) of the head of live nematodes under isotonic salt conditions (see **a**), hypertonic salt conditions (250 Osm, **b**), and hypertonic salt conditions (500 Osm, **c**). All the data was acquired from a depth of 25 μm inside the worms' head. The arrow lengths in the worm schematic illustrate the rate of water entrance to and exit from the nematode. The pixel dwell time and the frequency acquisition range were as in Fig. 1e. Images contain 100 × 200 pixels, resulting in a total image recording time of 400 s. Scale bar, 20 μm. (**d**) Average Brillouin shift of the pharynx (P, purple bar) and the surroundings (S, cyan bar) of live young adults ($N = 10$, $p$-value between levels of osmolarity $< 0.05$ as computed by one-way ANOVA following post-hoc analysis Tukey's tests, $p$-value within levels of osmolarity $< 0.05$, as computed by paired student's $t$-test, 25-μm depth) before and after hyperosmotic shock of 125 mM NaCl (250 Osm) and 250 mM NaCl (500 Osm). Error bars represent standard deviation from the mean. (**e**) Average Brillouin linewidth of the nematodes' pharynx and surroundings before and after the hyperosmotic shock (same statistics as in **d**). These findings suggest that the gigahertz-frequency stiffness and viscosity of the worm's pharyngeal region increases under hyperosmotic stress, whereas the pharynx is stiffer and more viscous than its surroundings regardless of the osmotic conditions.

Another application of SBS microscopy is biomechanical mesoscopy. Although spontaneous Brillouin microscopy has been used to acquire mesoscopic images of the Brillouin shift of tissue[11], mouse embryo[15], and zebrafish[18], the recording periods are often excessively long due to prolong pixel dwell times (hundreds of milliseconds). While shorter times (50 ms) are possible for imaging thin biological cells with spontaneous Brillouin microscopes based on virtually imaged phase arrays[20], their sub gigahertz spectrometer resolution considerably limits the ability to detect biomechanical constituents with highly overlapping Brillouin bands[17]. We map here in vivo the mesoscopic distributions of the Brillouin shift and the Brillouin linewidth across the middle plane of *C. elegans* nematodes at three larval stages (L2, L3, L4) and two adult stages (young adult and adult), as shown in Fig. 4a,b. These distribution maps allow us to identify organs and structures, such as the pharynx, vulva, gonad, and eggs, within the worms during development based on their viscoelastic contrast. Focusing on the nematode's head, a noticeable feature in these maps is the larger Brillouin shifts and linewidths of the pharynx than those of its surroundings, at all developmental stages. This was further validated statistically in several nematodes by means of the average Brillouin shift and linewidth of the pharynx (inset of Fig. 4a) and the surrounding region (inset of Fig. 4b). These findings imply that, regardless of the particular developmental stage of the nematode, the pharynx is stiffer and more viscous than the surrounding area at high frequencies. Interestingly, the mean Brillouin shift and linewidth of the pharynx and surroundings remained relatively unchanged during development. This suggests that the mechanical properties of the pharynx and the surrounding area are determined early in development, perhaps because of the prominent function of the pharynx in feeding[33].

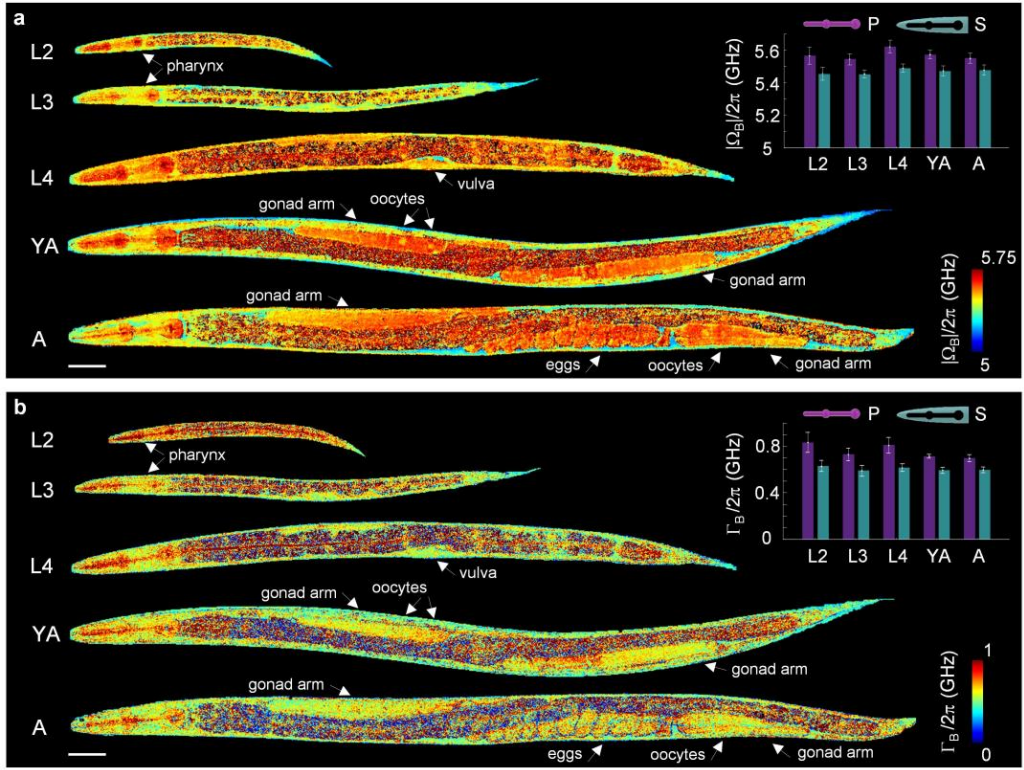

**Fig. 4.** Biomechanical SBS mesoscopy of developing *C. elegans* nematodes. (**a**) Brillouin shift $\Omega_B$ images of live nematodes at three larval stages (L2, L3, L4) and two adult stages, young adult (YA) and adult (A). The inset shows the average Brillouin shift of the pharynx (P, purple bar) and the surroundings (S, cyan bar) of the live worms at the different stages ($N = 10$ per stage, *p*-value within stages $< 0.05$ as computed by paired student's *t*-test, *p*-value for the pharynx between the L4 stage and the other stages $< 0.05$, all other *p*-values between stages were $> 0.05$ as computed by one-way ANOVA following post-hoc analysis Tukey's tests). Error bars represent standard deviation from the mean. Images were acquired from the middle plane of the nematodes at depths of 10-35 μm inside the worms. The pixel dwell time and the frequency acquisition range were as in Fig. 1e. Images contain $100 \times 400 – 120 \times 1200$ pixels, resulting in a total image recording time of 800 – 2880 s. Scale bar, 50 μm. (**b**) Brillouin linewidth $\Gamma_B$ images of the live nematodes at the different developmental stages. The average Brillouin linewidth of the worms' pharynx and surroundings is displayed in the inset for the various stages (same statistics as in **a**). SBS microscopy enables in vivo mechanical contrast visualization across large areas inside live organisms.

With its high sensitivity and spectrometer resolution, SBS microscopy enables mechanical imaging of large and living samples in practical times and the ability to highly specify mechanical constituents inside the samples based on resolvable Brillouin information. Together, these capabilities may open up new possibilities for studying various mechanical aspects of cancer and muscle disorders in model organisms.

**Methods**

**1. SBS microscope**

A detailed schematic of the SBS microscope is shown in Supplementary Fig. 1a. An amplified continuous-wave (CW) distributed feedback (DFB) laser (SYST TA-pro-DFB, Toptica) and a CW DFB laser (SYST DL-100-DFB, Toptica), both s-polarized, thermally stabilized and coupled into a polarization-maintaining single-mode fiber, serve as a pump beam at a frequency $\omega_1$ and a probe beam at a frequency $\omega_2$, respectively. The frequency $\omega_1$ corresponds to the wavelength of 780.24 nm, and the frequency $\omega_2$ is scanned around the Stokes frequency $\omega_1 - \Omega_B$, with $\Omega_B$ representing the characteristic longitudinal Brillouin shift of the material. Although SBG and SBL spectra provide equivalent information, we acquired SBG spectra because the passband of the detection filter used is flatter for the Stokes than for the anti-Stokes frequency band[26]. The lasers exhibit mode-hop-free tuning ranges larger than 50 GHz and laser linewidths of 1-4 MHz, providing wide spectral range and high spectrometer resolution for the acquisition of SBG spectra. The frequency of the probe laser is scanned by modulating its current with a sawtooth wave from a function generator (AFG2021, Tektronix). The frequency difference between the pump and probe lasers $\Delta\omega = |\omega_2 - \omega_1|$ is determined from a lookup table produced by measuring the beat frequency of the two lasers as a function of the probe laser current with a fast photodetector (1434, Newport) and a frequency counter (EIP 578B, Phase Matrix). The pump beam is optically filtered by a narrowband Bragg filter (SPC-780, OptiGrate) to suppress the amplified spontaneous emission of the pump laser. The filtered beam is modulated with an acousto-optic modulator (15210, Gooch and Housego) driven by a 1.1 MHz sinusoidal wave from the function generator. The same sinusoidal wave is used as a reference for the lock-in amplifier. Following modulation, the pump beam is p-polarized by a half wave plate and is transmitted toward the sample through a polarizing beam splitter. The pump and probe beams are circularly polarized orthogonally to each other with quarter wave plates, and are tightly focused at a joint point in the sample by aspheric lenses with a numerical aperture (NA) of 0.25 (Asphericon). This polarization scheme allows an efficient modulation of the pump beam and also SBS in a backscattering geometry, which maximizes the obtainable SBS signal. As a nonlinear process, SBS yielded with this NA an estimated lateral resolution of $\lambda / (4 \times NA) = 0.78$ μm and an estimated axial resolution of $\lambda \times n / NA^2 = 16.6$ μm in water. The sample is mounted on a close-loop XYZ motorized stage (LS-50, Applied Scientific Instrumentation) to enable raster scanning of the sample through the joint focus of the aspheric lens pair. The backscattered SBG signal at the probe frequency $\omega_2$ is collected with the same aspheric lens that focuses pump light into the sample, is then s-polarized by the quarter wave plate of the pump beam, and is finally reflected toward the detection module through the polarizing beam splitter. The detection unit comprises an atomic vapor notch filter at the pump frequency $\omega_1$, an amplified photodetector, and a lock-in amplifier. The atomic filter is based on the narrowband absorption of a rubidium-85 cell (SC-RB85-25x150-Q-AR, Photonics Technologies), heated to 90°C, at the Fg=3 line frequency to which the pump laser is tuned for suppressing unwanted pump stray reflections in the lock-in detection bandwidth. The amplified photodetector uses a large-area photodiode (FDS1010, Thorlabs) reverse-biased at 24V and a novel transimpedance amplifier with gain and equivalent input noise of $5 \times 10^4$ V/A and 2.5 pA/√Hz at the lock-in modulation frequency of 1.1 MHz, respectively, to generate an ac-coupled output voltage proportional to the SBG. In addition, the amplified photodetector has a dc-coupled voltage output to monitor the probe intensity attenuated by transmission through the sample. A high-frequency lock-in amplifier (SR844, Stanford Research) is used to demodulate the probe intensity. A data acquisition unit (NI USB-6212 BNC, National Instruments) connected to a personal computer is fed with the analog x-output of the lock-in amplifier, the dc-coupled voltage output of the amplified photodetector, the sawtooth wave output of the function generator, and the analog readout channel of the frequency counter.

To acquire a SBG spectrum, the lock-in detection bandwidth is set in the lock-in amplifier and the sawtooth wave is programmed in the function generator, with the height of the waveform defining the scanning frequency range of the probe laser around the Stokes frequency $\omega_1 - \Omega_B$ and the time period of the waveform determining the acquisition time of the entire spectrum. The microscope is focused into a point in the sample and a custom program (LabVIEW version 2014, National Instruments) tunes the frequency of the probe laser and records the dc and ac intensity of the transmitted probe beam for each $\Delta\omega$. Correction of the SBG signal for sample attenuation is performed by estimating the attenuation coefficient of the sample and the pump intensity at the focusing point from the measured dc probe intensity and the predetermined imaging depth in the sample. Spectral analysis was performed in MATLAB (The MathWorks Inc.) to fit a measured SBG spectrum to a double lorentzian line (Fig. 1d).

For biomechanical SBS imaging, the sample is raster-scanned through the microscope focus with a XYZ motorized stage controlled by a multi-axis controller (MS-2000, Applied Scientific Instrumentation). The beginning of a line scan is initiated by the controller, which also triggers, via a transistor-transistor logic (TTL) signal, the function generator that drives the current of the probe laser for frequency scanning (Supplementary Fig. 1b). The period of the driving current defines the pixel dwell time over which a SBG spectrum is acquired with predetermined measurement bandwidth and pixel size (Supplementary Fig. 1b). The pixel size (or the scanning step size) should be at least two times smaller than the optical resolution for Nyquist sampling. We set the pixel size to closely match the optical resolution (1 μm), which represents a compromise between the acquisition speed and the spatial resolution of the image. At each pixel, a SBG spectrum is acquired by the lock-in amplifier, as described above (Supplementary Fig. 1b). When the motorized stage completes the line scan the TTL signal falls, which leads to termination of frequency scanning and movement of the stage to the subsequent line scan, as shown in Supplementary Fig. 1b.

For brightfield imaging, a halogen light source (HL-2000, Ocean Optics) was introduced into the SBS microscope by placing a dichroic mirror in the path of the probe beam, and a complementary metal–oxide–semiconductor camera (Lt225, Lumenera) was imaging the sample through a dichroic mirror located immediately before the detection unit of the microscope, as shown in Supplementary Fig. 1a.

## 2. Detection limit of the SBS microscope

We previously have demonstrated an SBS spectrometer for high-speed analysis of materials[26–28]. Here, we use a novel low-noise high-frequency transimpedance photoreceiver to yield shot-noise limited sensitivity of the spectrometer. The electrical noise of our lock-in amplifier is 1.8 nV/√Hz, therefore dominating over the shot noise at detected optical powers of 0.1-10 mW. This noise performance is insufficient for SBS bioimaging with high sensitivity. An efficient technique to improve the sensitivity is to increase the signal-to-noise ratio (SNR) using a low-noise transimpedance photoreceiver. To this end, a large-area photodiode (FDS1010, Thorlabs) was integrated with a custom transimpedance amplifier with gain and equivalent input noise of $5\times10^4$ V/A and 2.5 pA/$\sqrt{Hz}$ at the lock-in modulation frequency of 1.1 MHz, respectively. Using this low-noise high-frequency transimpedance photoreceiver, the SBS microscope was brought close to the shot-noise limit over the entire range of optical powers of 0.1-10 mW (Supplementary Fig. 2a). Further, we doubled the SBG signal detected by improving the crossing efficiency of the pump and probe beams in the sample to 55% while minimizing the amount of pump stray reflections detected. This was accomplished using measurement chambers that are much thicker than the confocal parameter of the focusing lenses. Consequently, we were able to measure SBG spectra of water in acquisition times as low as 2 ms with SNRs > 25 dB, resulting in Brillouin shift and linewidth measurement precision of $\delta\Omega_B/2\pi \leq 11.5$ MHz and $\delta\Gamma_B/2\pi \leq 35$ MHz, respectively (Supplementary Fig. 2b-d). This result represents a fivefold improvement in the spectral acquisition time compared with the previous SBS spectrometer, without compromising precision of the measurements. For the mean Brillouin shift and linewidth retrieved from the SBG spectra of water measured in 2 ms ($\Omega_B/2\pi = 5.05$ GHz, $\Gamma_B/2\pi = 323$ MHz), the SBS microscope obtained fractional precision of at least $2 \times \delta\Omega_B/\Omega_B = 0.0046$ and $[(\delta\Gamma_B/\Gamma_B)^2 + (\delta\Omega_B/\Omega_B)^2]^{1/2} = 0.11$ in the estimate of the real and imaginary parts of the gigahertz-frequency longitudinal modulus $M^*$ of water, respectively.

## 3. Preparation of the *C. elegans* samples

Wild type (N2) *C. elegans* worms were grown on nematode growth media (NGM) plates seeded with the Escherichia coli OP50 or OP50-1 strains at 15°C. 30-60 embryos, laid at 15°C, were picked, transferred to new plates and grown at 25°C for the duration of the experiment. Nematodes at a well-defined developmental stage were determined using a light microscope during the developmental time window of interest. During the reproductive period, animals were transferred to fresh plates every 1-2 days to circumvent progeny contamination.

For imaging, we first prepared two 0.5-mm-thick 5% agar pads mixed with 10 mM sodium azide solution (NaN3) to anesthetize the worms. Then, the agar pads were mounted on two 0.17-mm-thick round glass coverslips (18 mm and 25 mm in diameter), and 10-15 nematodes at a specific developmental stage were sandwiched between the agar-padded coverslips (Supplementary Fig. 3a). 10 µl of M9 contact buffer was added between the agar pads and served as a control environment for the worms. To fix the entire sample and to avoid dehydration, a few drops of UV glue were applied at the edge of the smaller coverslip and a thin layer of Vaseline sealed the gap between the two coverslips, as illustrated in Supplementary Fig. 3b. Following SBS imaging, worms were washed from the agar pads to NGM plates for recovery.

## 4. Measurements of photodamage in SBS imaging of live *C. elegans* nematodes

In live-organism SBS imaging, high irradiation intensities are used. To assess photodamage, we monitored the Brillouin shift at multiple locations in the head of live *C. elegans* nematodes and across the agar pads surrounding the worm over 120 s at room temperature (20°C) (Supplementary Fig. 4). These measurements showed no consistent elevation in the Brillouin shift with time.

Because the Brillouin shift increases with rising temperature[39], these results suggest that no substantial sample heating occurred in 120 s – a time period significantly longer than the pixel dwell time of 20 ms used in SBS imaging of live *C. elegans* worms. In addition, the maximum standard deviation of these Brillouin shift measurements is 0.021 MHz, corresponding to a maximum temperature change of less than 1.8°C in water[40]. This result is consistent with our calculation of the heating in water (below 1 K) due to water absorption at 780 nm using the same conditions as in the experiments (total excitation power of 265 mW, NA of 0.25, and measurement time of 120 s). In addition, examination of several worms following SBS imaging showed recovery from anesthesia and a typical crawling motion.

## 5. Estimate of the mass density and the complex longitudinal modulus of materials by SBS

In SBS, the mean density of a material $\rho$ is related to the Brillouin shift $\Omega_B$ and linewidth $\Gamma_B$, and the normalized line-center gain factor $g_B^N$ of the SBG spectrum of the medium by the expression

$$\rho = \frac{\gamma_e^2 \omega_2 q_B^2}{2n^2 c^2 \Omega_B \Gamma_B g_B^N}, \tag{1}$$

where $\gamma_e$ is the electrostrictive constant, $\omega_2$ is the frequency of the probe beam, $n$ is the refractive index of the medium, $c$ is the speed of light in vacuum, and $q_B$ is the acoustic wavenumber. $g_B^N$ is calculated from the SBG spectrum $G(\Omega)$ as

$$g_B^N = \frac{G_B}{\eta L_{eq} I_1}, \tag{2}$$

with $G_B$ being the peak gain $G(\Omega_B)$, $\eta$ the crossing efficiency of the pump and probe beams in the sample, $I_1$ the intensity of the pump beam at the sample entrance, and $L_{eq}$ the equivalent SBS interaction length in the medium calculated using the assumed spatial distribution of the pump intensity in the sample[28]. For backward SBS, $q_B = 2 \times \omega_2 \times n / c$ and Eq. (1) reduces to[23]

$$\rho = \frac{2\gamma_e^2 \omega_2^3}{c^4 \Omega_B \Gamma_B g_B^N}. \tag{3}$$

By estimating $\gamma_e$ through use of the Lorentz–Lorenz law, it is obtained that $\gamma_e = (n^2 - 1) \times (n^2 + 2) / 3$ (Ref. 23). Thus, the mean density of a material $\rho$ can be evaluated from the measured values of $\Omega_B$, $\Gamma_B$, and $G_B$ of the SBG spectrum of the material and from knowledge about the material refractive index $n$. Subsequently, the complex longitudinal modulus of the material $M^*$ is calculated using the spectral parameters $\Omega_B$ and $\Gamma_B$ of the measured SBG spectrum, the $\rho$ estimated by Eq. (3), and the literature refractive index value of the material $n$ as

$$M^* = \rho \frac{\lambda_1^2}{4n^2} \Omega_B^2 \cdot \left(1 + i \frac{\Gamma_B}{\Omega_B}\right), \tag{4}$$

with $\lambda_1$ denoting the wavelength of the beam at $\omega_1$ (Refs. 7 and 30).

We validated Eq. (3) and computed $M^*$ (Eq. (4)) in various materials using the SBS microscope with 0.033 NA focusing lenses (Supplementary Fig. 5). The use of these low NA lenses minimized the effects of broadening and blue shift of the SBG spectrum and enabled to work in a near backward SBS geometry with a crossing efficiency close to 100%. Reported refractive index values at a wavelength of 780.24 nm and at 20°C were used in Eqs. (3) and (4) for the materials measured[41,42]. The estimated mass density of all materials agreed well, to within 1-10%, with values from the literature (Ref. 43; Supplementary Table I).

## 6. Estimate of the mass density and the complex longitudinal modulus in *C. elegans* by the SBS microscope

In backward SBS, the SBG spectrum of the $i^{\text{th}}$ voxel of volume $\Delta x \times \Delta y \times \Delta z$ centered at $(x_i, y_i, z_i)$ in a scattering medium is[25]

$$G(\Omega) = \eta g(\Omega) f_{eq}\left(L^i_{eff}\right) I^i_1, \qquad (5)$$

where $\eta$ the crossing efficiency of the pump and probe beams in the sample, $g(\Omega)$ is the SBS gain factor which, to good approximation, is described by a lorentzian with a spectral shift $\Omega_B$, linewidth $\Gamma_B$, and peak gain $G_B$. $L^i_{eff}$ is the effective SBS interaction length in the medium

$$L^i_{eff} = \bar{\mu}_t^{-1}(x_i, y_i) \cdot \left(1 - e^{-\bar{\mu}_t(x_i, y_i)\Delta z}\right), \qquad (6)$$

with $\bar{\mu}_t(x_i, y_i)$ being the mean total attenuation coefficient of the medium at $(x_i, y_i)$. $f_{eq}(\cdot)$ is a function that describes the dependence of the SBS interaction length on the spatial distribution of the pump intensity in the sample[28]. $I^i_1$ is the pump intensity at the voxel entrance

$$I^i_1 = I_{10} e^{-\bar{\mu}_t(x_i, y_i)(z_i - \Delta z/2)}, \qquad (7)$$

where $I_{10}$ is the pump intensity at the entrance plane of the medium.

To estimate the mass density of the *C. elegans* nematode at the $i^{\text{th}}$ voxel, we used Eq. (3) with the reported refractive index $n$ value of the region to which the voxel belongs in the worm (34), the Brillouin shift $\Omega_B$ and linewidth $\Gamma_B$ retrieved from the SBG spectrum $G(\Omega)$ measured at the voxel, and the normalized Brillouin line-center gain factor $g_B^{N,i}$ evaluated as

$$g_B^{N,i} = \frac{G(\Omega_B)}{\eta f_{eq}\left(L^i_{eff}\right) I^i_1}. \qquad (8)$$

Here, the mean total attenuation coefficient $\bar{\mu}_t(x_i, y_i)$, appearing in $L^i_{eff}$ (Eq. (6)) and $I^i_1$ (Eq. (7)), was recovered by measuring the attenuation of the probe beam through the worm at $(x_i, y_i)$ using the dc output signal of the photoreceiver, together with estimating the thickness of the nematode at the same point from the worm's brightfield images (assuming cylindrical symmetry of the worm's body). The voxel's axial size $\Delta z$ in Eqs. (6) and (7) was approximated as the extent of the resolution cell along the axial dimension, and the voxel's axial position $z_i$ in Eq. (7) was obtained from the position readings of the motorized sample stage.

To compute the complex longitudinal modulus of the material $M^*$ at the $i^{th}$ voxel of the *C. elegance* nematode, Eq. (4) is used with the spectral parameters $\Omega_B$ and $\Gamma_B$ of the SBG spectrum measured at the $i^{th}$ voxel, the mass of density $\rho$ estimated at the voxel, and the literature refractive index $n$ value of the region to which the voxel belongs in the worm.

We point out that Eq. (5) assumes low NA values (< ~0.1). For the NA of 0.25 used in SBS imaging of the *C. elegance* worms, the Brillouin shift is expected to show a <1% decrease compared with the Brillouin shift at NA = 0.1, the Brillouin linewidth a <5% increase, and the normalized Brillouin line-center gain factor a <4% decrease, resulting in a <2% (<4%) decrease (increase) in the calculation of the real (imaginary) part of the complex longitudinal modulus.

## References

1. Thomas, G., Burnham, N. A., Camesano, T. A. & Wen, Q. Measuring the mechanical properties of living cells using atomic force microscopy. *J. Vis. Exp.* **76**, 50497 (2013).
2. Essmann, C. L. et al. In-vivo high resolution AFM topographic imaging of Caenorhabditis elegans reveals previously unreported surface structures of cuticle mutants. *Nanomedicine* **13**, 183-189 (2017).
3. Goulam Houssen, Y., Gusachenko, I., Schanne-Klein, M. C. & Allain, J. M. Monitoring micrometer-scale collagen organization in rat-tail tendon upon mechanical strain using second harmonic microscopy. *J. Biomech.* **44**, 2047–2052 (2011).
4. Wang, S. & Larin, K. V. Shear wave imaging optical coherence tomography (SWI-OCT) for ocular tissue biomechanics. *Opt. Lett.* **39**, 41-44 (2014).
5. Kennedy, K. M. et al. Quantitative micro-elastography: imaging of tissue elasticity using compression optical coherence elastography. *Sci. Rep.* **5**, 15538 (2015).
6. Kennedy, B. F., Wijesinghe, P. & Sampson, D. D. The emergence of optical elastography in biomedicine. *Nat. Photonics* **11**, 215–221 (2017).
7. Vaughan, J. M. & Randall, J. T. Brillouin scattering, density and elastic properties of the lens and cornea of the eye. *Nature* **284**, 489-491 (1980).
8. Koski, K. J., Akhenblit, P., McKiernan, K. & Yarger, J. L. Non-invasive determination of the complete elastic moduli of spider silks. *Nat. Mater.* **12**, 262-267 (2013).
9. Palombo, F. et al. Biomechanics of fibrous proteins of the extracellular matrix studied by Brillouin scattering, *J. R. Soc. Interface* **11**, 20140739 (2014).
10. Scarcelli, G. et al. Noncontact three-dimensional mapping of intracellular hydromechanical properties by Brillouin microscopy. *Nat. Methods* **12**, 1132-1134 (2015).
11. Antonacci, G. et al. Quantification of plaque stiffness by Brillouin microscopy in experimental thin cap fibroatheroma. *J. R. Soc. Interface* **12**, 20150483 (2015).
12. Lepert, G., Gouveia, R. M., Connonb, C. J. & Paterson, C. Assessing corneal biomechanics with Brillouin spectro-microscopy. *Faraday Discuss.* **187**, 415-428 (2016).
13. Elsayad, K. et al. Mapping the subcellular mechanical properties of live cells in tissues with fluorescence emission-Brillouin imaging. *Sci. Signal.* **9**, rs5 (2016).
14. Meng, Z., Traverso, A. J., Ballmann, C. W., Troyanova-Wood, M. A. & Yakovlev, V. V. Seeing cells in a new light: a renaissance of Brillouin spectroscopy. *Adv. Opt. Photonics* **8**, 300-327 (2016).
15. Raghunathan, R. et al. Evaluating biomechanical properties of murine embryos using Brillouin microscopy and optical coherence tomography. *J. Biomed. Opt.* **22**, 086013 (2017).

16. Karampatzakis, A. et al. Probing the internal micromechanical properties of Pseudomonas aeruginosa biofilms by Brillouin imaging. *NPJ Biofilms Micro.* **3**, 20 (2017).
17. Mattana, S. et al. Non-contact mechanical and chemical analysis of single living cells by microspectroscopic techniques. *Light Sci. Appl.* **7**, 17139 (2018).
18. Schlußler, R. et al. Mechanical mapping of spinal cord growth and repair in living zebrafish larvae by Brillouin imaging. *Biophys. J.* **115**, 911-923 (2018).
19. Bevilacqua, C., Sánchez-Iranzo, H., Richter, D., Diz-Muñoz, A. & Prevedel, R. Imaging mechanical properties of sub-micron ECM in live zebrafish using Brillouin microscopy. *Biomed. Opt. Express* **10**, 1420-1431 (2019).
20. Nikolić, M. & Scarcelli, G., Long-term Brillouin imaging of live cells with reduced absorption-mediated damage at 660nm wavelength. *Biomed. Opt. Express* **10**, 1567-1580 (2019).
21. Chiao, R. Y., Townes, C. H. & Stoicheff, B. P. Stimulated Brillouin scattering and coherent generation of intense hypersonic waves. *Phys. Rev. Lett.* **12**, 592-595 (1964).
22. Damzen, M. j. *Stimulated Brillouin Scattering: Fundamentals and Applications* (IOP Publishing, Bristol, 2003).
23. Boyd, R. W. *Nonlinear Optics* (Academic Press, New York, ed. 3, 2008).
24. Ballmann, C. W. et al. Stimulated Brillouin scattering microscopic imaging. *Sci. Rep.* **5**, 18139 (2015).
25. Remer, I. & Bilenca, A. Background-free Brillouin spectroscopy in scattering media at 780 nm via stimulated Brillouin scattering. *Opt. Lett.* **41**, 926-929 (2016)
26. Remer, I. & Bilenca, A. High-speed stimulated Brillouin scattering spectroscopy at 780 nm. *APL Photonics* **1**, 061301 (2016).
27. Remer, I., Cohen, L. & Bilenca, A. High-speed continuous-wave stimulated Brillouin scattering spectrometer for material analysis. *J. Vis. Exp.* **127**, e55527 (2017).
28. Remer, I. Stimulated Brillouin Scattering Microscopy. (Ben-Gurion University of the Negev, 2017).
29. Ballmann, C. W., Meng, Z., Traverso, A. J., Scully, M. O. & Yakovlev, V. V., Impulsive Brillouin microscopy. *Optica.* **4**, 124-128 (2017).
30. Litovitz, T. A. & Davis, C. M. *Physical Acoustics*, (Academic, New York, 1965, Vol. 2A)
31. Freudiger, C. W. et al. Label-free biomedical imaging with high sensitivity by stimulated Raman scattering microscopy. *Science* **322**, 1857-1861 (2008).
32. Kaletta, T. & Hengartner, M. O. Finding function in novel targets: C. elegans as a model organism. *Nat. Rev. Drug Discov.* **5**, 387-399 (2006).
33. Mango, S. E. The C. elegans pharynx: a model for organogenesis. *WormBook*, 1-26 (2007). 10.1895/wormbook.1.129.1.
34. Choi, W. et al. Tomographic phase microscopy. *Nat. Methods* **4**, 717–719 (2007).
35. Park, S. J., Goodman, M. B. & Pruitt, B. L. Analysis of nematode mechanics by piezoresistive displacement clam. *Proc. Natl. Acad. Sci. U S A.* **104**, 17376-17381 (2007).
36. Sznitman, J., Purohit, P. K., Krajacic, P., Lamitina, T. & Arratia, P. E. Material properties of Caenorhabditis elegans swimming at low Reynolds number. *Biophys. J.* **98**, 617-26 (2010).
37. Backholm, M., Ryu, W. S. & Dalnoki-Veress, K. Viscoelastic properties of the nematode Caenorhabditis elegans, a self-similar, shear-thinning worm. *Proc. Natl. Acad. Sci U S A.* **110**, 4528- 4533 (2013).
38. Gilpin, W., Uppaluri, S. & Brangwynne, C. P. Worms under pressure: bulk mechanical properties of *C. elegans* are Independent of the cuticle. *Biophys J.* **108**, 1887-1898 (2015).
39. Fry, E. S., Emery, Y., Quan, X. & Katz, J. W. Accuracy limitations on Brillouin lidar measurements of temperature and sound speed in the ocean, *Appl. Opt.* **36**, 6887-6894 (1997).
40. Schonle, A. & Hell, S. W. Heating by absorption in the focus of an objective lens. *Opt. Lett.* **23**, 325-327 (1998).
41. Podeă, J., Procházka, O. & Medin, A. Studies on agaroses; 1. Specific refractive index increments in dimethyl sulfoxide and in water at various wavelengths and temperatures. *Polymer* **36**, 4967-4970 (1995).
42. Polyanskiy, M. N. Refractive index database. Available at: http://www.refractiveindex.info (2008-2019).
43. Weast, R. C. *Handbook of Chemistry and Physics* (CRC Press, 1987).

# Supplementary Information

# High Sensitivity and Specificity Biomechanical Imaging by Stimulated Brillouin Scattering Microscopy

Itay Remer[1,2], Netta Shemsesh[3], Anat Ben-Zvi[3] & Alberto Bilenca[1,4*]

[1]Biomedical Engineering Department, Ben-Gurion University of the Negev, 1 Ben Gurion Blvd, Be'er-Sheva 84105, Israel
[2]Agilent Research Laboratories, 94 Shlomo Shmeltzer Road, Petach Tikva 4970602, Israel
[3]Department of Life Sciences, Ben-Gurion University of the Negev, 1 Ben Gurion Blvd, Be'er Sheva 84105, Israel
[4]Ilse Katz Institute for Nanoscale Science and Technology, Ben-Gurion University of the Negev,
1 Ben Gurion Blvd, Be'er-Sheva 84105, Israel

e-mail: *bilenca@bgu.ac.il

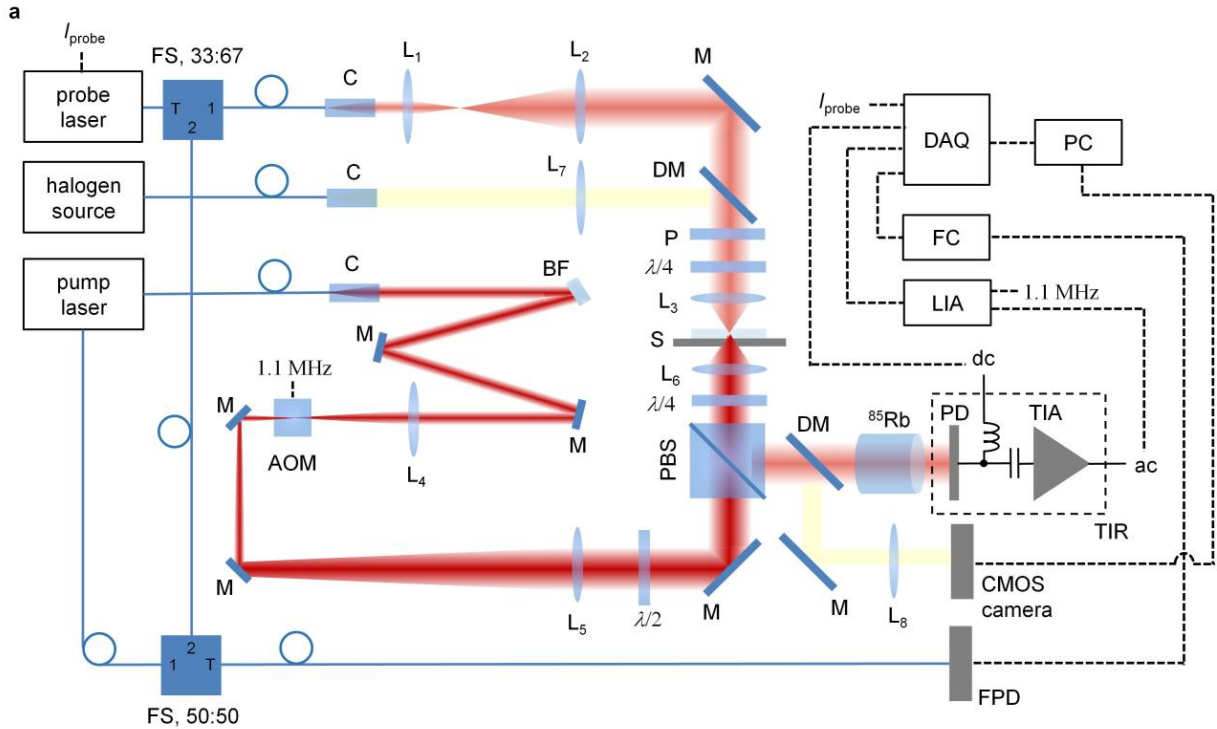

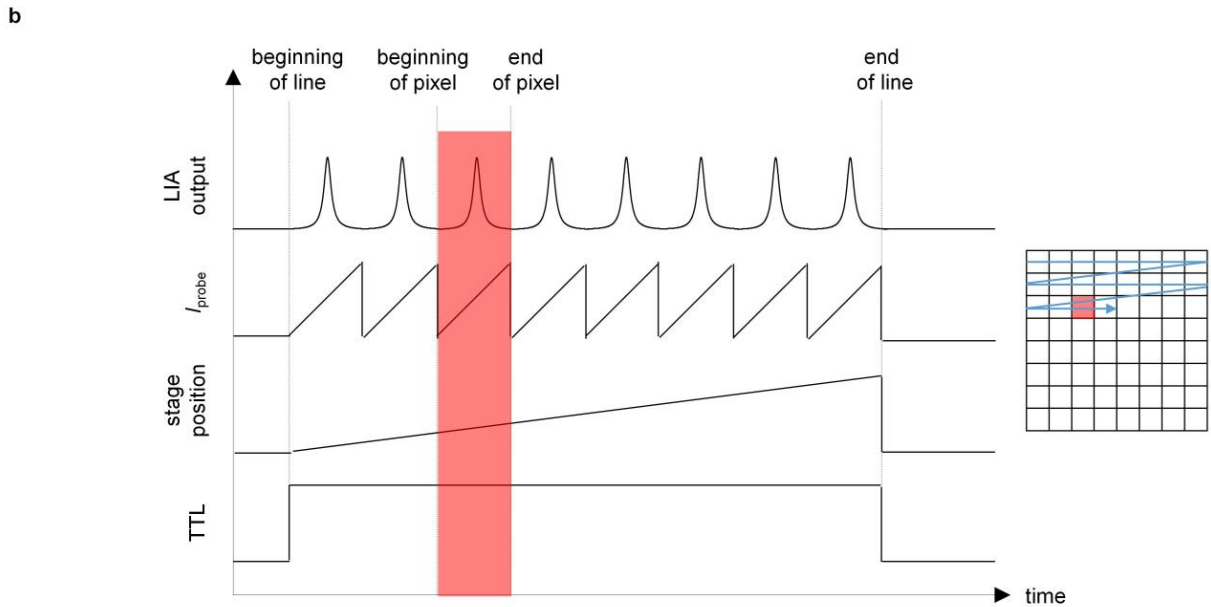

**Supplementary Fig. 1.** Detailed schematic of the SBS microscope. (**a**) Near infrared (NIR) fiber-coupled pump and probe distributed feedback (DFB) lasers are collimated with collimators of 1 mm (C). The collimated, s-polarized probe laser beam (light red) is expanded by lenses $L_1$ and $L_2$, and is then right circularly polarized by a quarter wave plate ($\lambda/4$) and focused into the sample (S) by a 0.25 NA aspheric lens $L_3$. The collimated, s-polarized pump laser beam (deep red) is filtered with a Bragg filter (BF) and is then expanded by lenses $L_4$ and $L_5$ and sinusoidally modulated by an acousto-optic modulator (AOM) at 1.1 MHz. Next, the beam is p-polarized by a half wave plate ($\lambda/2$) and transmitted through a polarizing beam splitter (PBS). The beam is subsequently left circularly polarized by a quarter wave plate ($\lambda/4$) and focused by an aspheric lens $L_6$ (same as lens $L_3$) to the focal point of lens $L_3$ in the sample. The backscattered SBG signal that modulates the intensity of the probe beam is collected with lens $L_6$ and is then s-polarized by a quarter wave plate ($\lambda/4$) and reflected by the PBS. Next, the signal passes a rubidium notch filter ($^{85}$Rb) at the pump frequency, and is detected by a transimpedance receiver (TIR), which comprises a large area photodiode (PD), a bias tee, and a transimpedance amplifier (TIA). The SBG signal in the ac output voltage of the TIR is measured by a lock-in amplifier (LIA). Two fiber splitters (FS) extract auxiliary beams from the pump and probe lasers, and a fast photodetector (FPD) and a frequency counter (FC) measure the beat frequency (or the difference frequency) between these beams. The SBG signal, the average probe power (measured by the dc output voltage of the TIR), the difference frequency measurement, and the probe laser current $I_{probe}$ are all sampled by a data acquisition card (DAQ) connected to a personal computer (PC) for further analysis and visualization. A wide halogen illumination and a CMOS camera co-register brightfield images in the microscope. The halogen beam is directed from a collimated fiber-coupled source to the sample through lenses $L_7$ and $L_3$ and a dichroic mirror (DM) that combines white and NIR light. The sample is imaged onto the camera via lenses $L_6$ and $L_8$, where a DM positioned after the PBS splits the NIR and white light for SBG and brightfield imaging, respectively. The camera is connected to a PC for further analysis and presentation. All folding mirrors (M) fit the microscope on a 18"×24" breadboard that is vertically mounted on an optical table. (**b**) SBS imaging acquisition scheme. The sample stage controller initiates the beginning of a line scan and TTL-triggers the probe current $I_{probe}$ to start the frequency scans. The motorized stage raster scans the sample through the joint focus of lenses $L_3$ and $L_6$, while the probe laser scans the pump-probe frequency difference around the Brillouin shift of the sample and the LIA retrieves the SBG spectrum at a pixel. The TTL signal falls when the stage completes the line scan and this terminates the frequency scans and moves the stage to the next line scan.

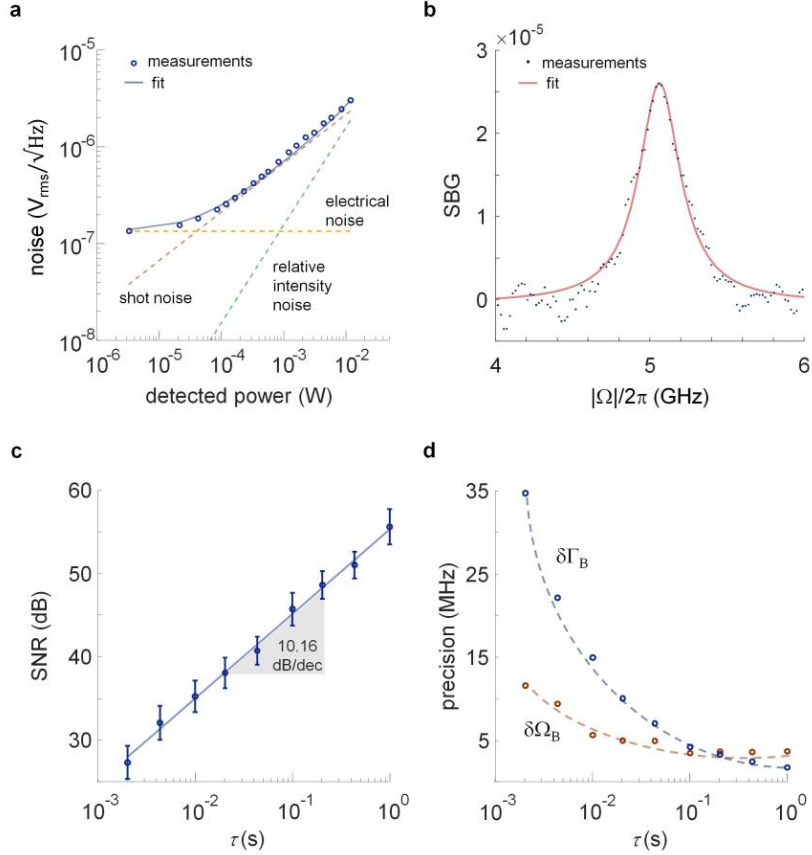

**Supplementary Fig. 2.** Performance of the SBS microscope. (**a**) Noise density against detected power. Measurements (open circles) were made in double-distilled water at room temperature (20°C) with only probe light excitation. To measure the noise density, the probe frequency $\omega_2$ was scanned around the Stokes frequency $\omega_1 - \Omega_B$ (~ −5.05 GHz from the pump frequency $\omega_1$) over a 2-GHz bandwidth in 100 ms and the ac signal at the receiver was lock-in detected at a 1.1-MHz reference frequency. The noise density was calculated as the ratio of the standard deviation of the ac voltage to the square-root of the lock-in bandwidth ($B$ = 138 Hz). The optical probe power was measured from the dc voltage at the receiver. The fit of the noise density measurements to the sum of three noise contributions is shown in a solid line. These noise contributions are the electrical noise density (125 nV/√Hz), the shot noise density ($2q \times R \times P$ where $q$ is the electron's charge, $R$ = 0.53 A/W is the photodetector responsivity, and $P$ is the optical power of the probe beam) and the relative intensity noise of the laser (RIN $\times R^2 \times P^2$ with RIN = 3.16 × 10$^{-17}$ Hz$^{-1}$). (**b**) SBG spectrum of double-distilled water at room temperature (20°C). The spectrum (close circles) was acquired over a 2-GHz bandwidth in 2 ms. A total excitation power of 265 mW was used. The lorentzian fit to the measurements is shown in a red solid line. (**c**) Signal-to-noise ratio (SNR) against the acquisition time of the SBG spectrum of water. The SNR was evaluated as the ratio of the SBG peak power to the standard deviation of the noise skirt in the SBG spectrum over a 1-GHz bandwidth. Open circles and error bars represent mean value and standard deviation from the mean, respectively, as calculated from one hundred SBG spectra measured sequentially. A slope of ~10 dB/dec was obtained from the linear regression line, as expected from the direct dependence of the SNR on the acquisition time (25). (**d**) Precision of the Brillouin shift and linewidth measurements ($\delta\Omega_B$ and $\delta\Gamma_B$) against the acquisition time of the SBG spectrum of water. The precision was calculated as the standard deviation of the Brillouin shift and linewidth recovered from the lorentzian fits of one hundred SBG spectra measured sequentially. Dashed lines are drawn to guide the eye.

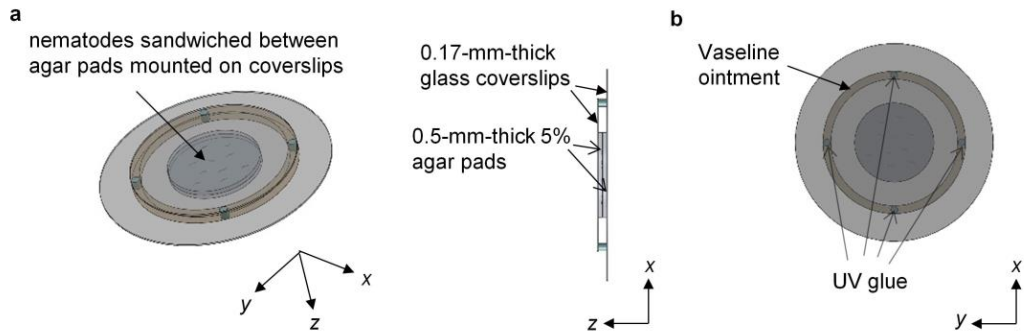

**Supplementary Fig. 3.** Sample mount for SBS imaging of live *C. elegans* nematodes. (**a**) Three-dimensional schematic and side view of the sample mount. Nematodes are sandwiched between 0.5-mm-thick agar pads each is mounted on a 0.17-mm-thick round glass coverslip of a different diameter (18 mm and 25 mm). (**b**) Top view of the sample mount. A few drops of UV glue are applied at the edge of the smaller coverslip and a thin layer of Vaseline seals the gap between the two coverslips, thereby fixing the entire sample and avoiding dehydration.

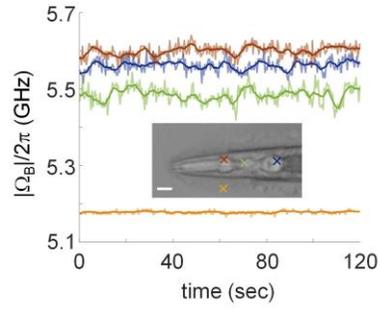

**Supplementary Fig. 4.** Photodamage in SBS imaging of live *C. elegans* nematodes. Brillouin shift $\Omega_B$ versus time for a live *C. elegans* worm. The Brillouin shift $\Omega_B$ was recovered from a double lorentzian fit of the SBG spectra measured over a 4-GHz bandwidth in 200 ms. For these measurements, a total excitation power of 265 mW was used, which is the same power level employed all through this work. Curve colors represent different spatial positions across the middle plane of the sample at which the measurements were taken (~25 μm in depth). These positions are marked in the coregistered brightfield image with crosses of corresponding colors (scale bar, 20 μm). Curve pale and dark shades indicate raw and ten-sample moving average data, respectively.

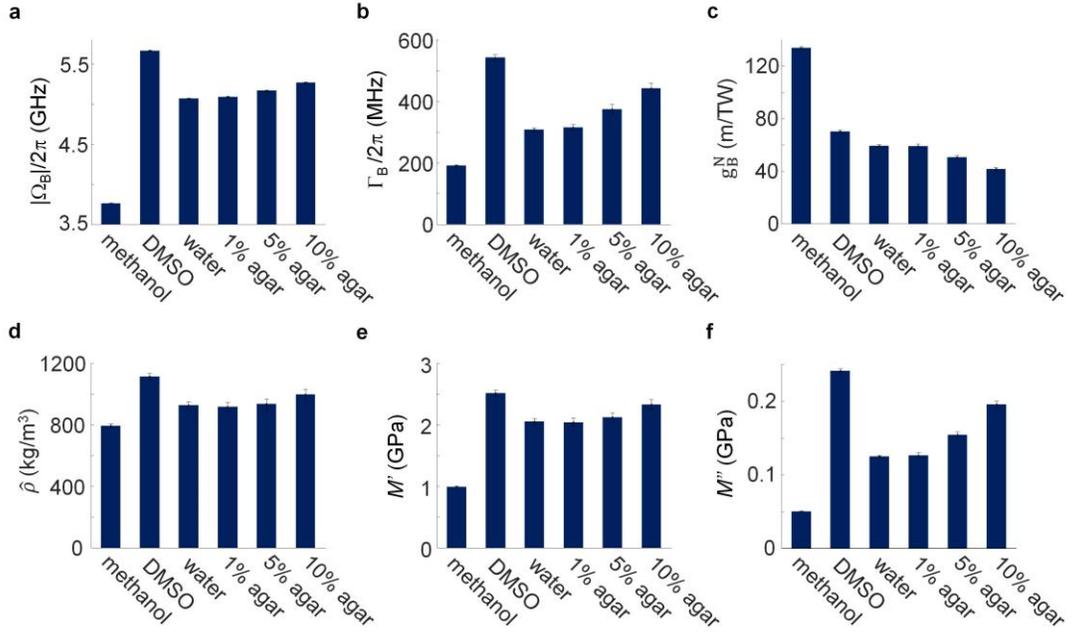

**Supplementary Fig. 5.** Estimation of the mass density and the complex longitudinal modulus of materials by SBS. (**a**) Brillouin shift $\Omega_B$ of methanol, dimethyl sulfoxide (DMSO), double-distilled water, and 1%, 5% and 10% agar/water gels at room temperature (20°C). (**b**) Brillouin linewidth $\Gamma_B$ of the different materials. (**c**) Normalized Brillouin line-center gain factor $g_B^N$ of the different materials. The spectral parameters ($\Omega_B$, $\Gamma_B$, $g_B^N$) were retrieved from the lorentzian fits of two hundred SBG spectra measured sequentially. Each spectrum was recorded over a bandwidth of 4 GHz in 200 ms by the SBS microscope with 0.033 NA focusing lenses and a total excitation power of 285 mW. (**d**) Mass density estimate for the different materials using Eq. (3) of Methods Section 5 with refractive index values of 1.315, 1.47, 1.329, 1.331, 1.337, and 1.345 for methanol, DMSO, double-distilled water, 1%, 5% and 10% agar/water gels, respectively. (**e**) Real part of the gigahertz-frequency longitudinal modulus $M'$ for the different materials. (**f**) Imaginary part of the gigahertz-frequency longitudinal modulus $M''$ for the different materials. $M'$ and $M''$ were calculated using Eq. (4) of Methods Section 5 with the Brillouin shift $\Omega_B$ and linewidth $\Gamma_B$ values described in **a** and **b**, and the mass density estimates and refractive indices presented in **d**.

|  | methanol | DMSO | water | 1% agar | 5% agar | 10% agar |
|---|---|---|---|---|---|---|
| $\|\Omega_B\|/2\pi$ (GHz) | 3.764±0.04 | 5.666±0.04 | 5.07±0.004 | 5.092±0.005 | 5.167±0.006 | 5.268±0.006 |
| $\Gamma_B/2\pi$ (MHz) | 191±3 | 543±9 | 308±6 | 316±10 | 375±14 | 442±17 |
| $g_B^N$ (m/TW) | 133.69±1.38 | 70.39±0.92 | 59.58±0.83 | 59.39±1.52 | 50.77±0.43 | 42.06±1.03 |
| $\hat{\rho}$ (kg/m³) | 794±15 | 1113±22 | 929±21 | 933±29 | 935±32 | 997±35 |
| $\rho$ (kg/m³) | 791 | 1100 | 1000 | 1009 | 1045 | 1090 |
| $M'$ (GPa) | 0.989±0.018 | 2.518±0.050 | 2.056±0.048 | 2.078±0.065 | 2.125±0.075 | 2.331±0.083 |
| $M''$ (GPa) | 0.0504±0.001 | 0.2413±0.003 | 0.1249±0.002 | 0.1267±0.003 | 0.1543±0.004 | 0.1957±0.005 |

**Supplementary Table I.** Mean and standard deviation values for the Brillouin shift $\Omega_B$, the Brillouin linewidth $\Gamma_B$, the normalized Brillouin line-center gain factor $g_B^N$, the mass density estimate $\hat{\rho}$, and the real and imaginary parts of the gigahertz-frequency longitudinal modulus $M'$ and $M''$ of methanol, DMSO, double-distilled water, and 1%, 5% and 10% agar/water gels at room temperature (20°C). Literature values for the mass density are also presented[43].